\documentclass[aps,pra,twocolumn,groupedaddress,amsmath,amssymb]{revtex4}
\usepackage{natbib}
\usepackage{graphicx}

\begin{document}

%
%

\title{EIT and diffusion of atomic coherence}

\author{I.\ Novikova$^*\dagger$\thanks{${}^{*}$Corresponding author. E-mail: inovikova@cfa.harvard.edu}
, Y.\
Xiao$\dagger$, D.\ F.\ Phillips$\dagger$, and R.\ L.\
Walsworth$\dagger\ddagger$\\
 $\dagger$ Harvard-Smithsonian Center for Astrophysics, Cambridge, MA, 02138, USA\\
 $\ddagger$  Department of Physics, Harvard University, Cambridge,
MA, 02138, USA
%
%
}
\date{\today}


\begin{abstract}
We study experimentally the effect of diffusion of Rb atoms  on
Electromagnetically Induced Transparency (EIT) in a buffer gas
vapor cell. In particular, we find that diffusion of atomic
coherence in-and-out of the laser beam plays a crucial role in
determining the EIT resonance lineshape and the stored light
lifetime.
\end{abstract}

\maketitle 

\begin{figure}
\includegraphics[width=1.0\columnwidth]{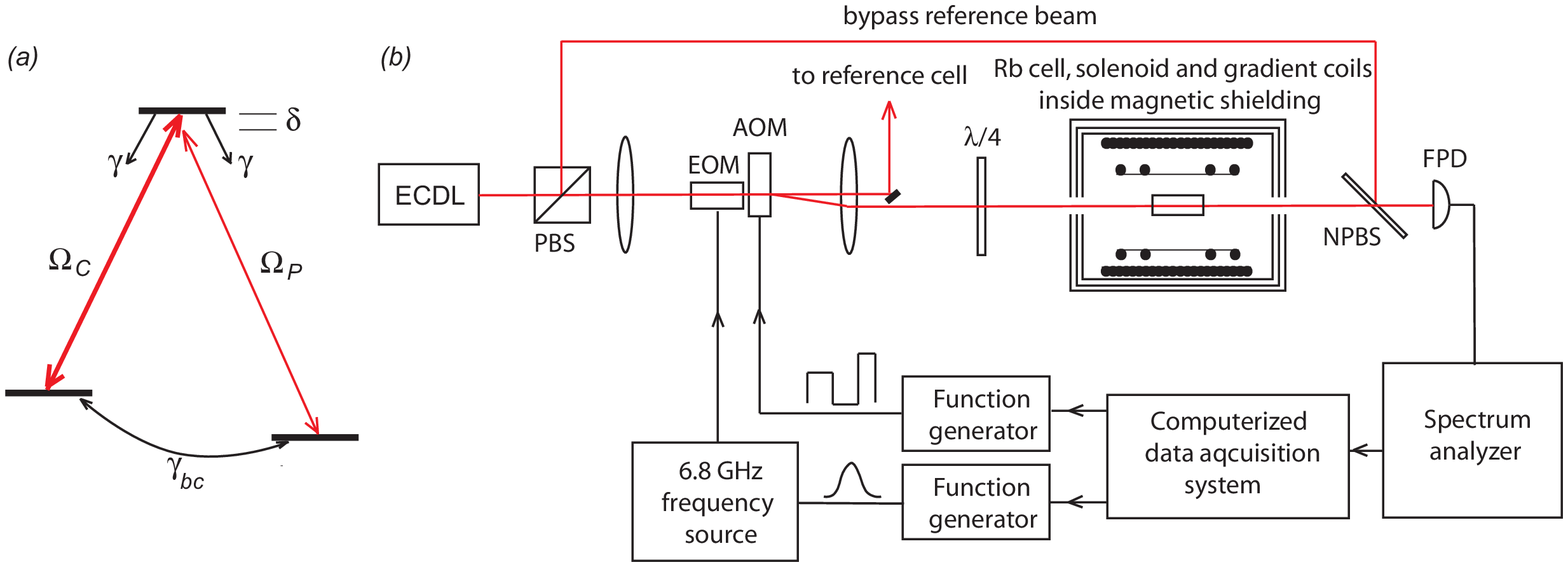}
\caption{ a) Simplified three-level $\Lambda$ system; b) Schematic
of the experimental setup (see text for abbreviations).}
\label{setup.fig}
\end{figure}

In this paper we address an important issue for
electromagnetically induced transparency
(EIT)~\cite{harris'97pt,scullybook} and stored
light~\cite{fleischhauer02pra,lukin03rmp} in alkali vapor cells
(i.e., warm atoms with buffer gas): the practical role of the
diffusion of atomic coherence in-and-out of the laser fields that
prepare and probe the atoms. EIT occurs in a three-level
$\Lambda$-system in which two coherent electromagnetic fields (the
probe and control fields) are in two-photon Raman resonance with
two atomic ground-state sublevels, as shown in
Fig.~\ref{setup.fig}(a). The atoms are optically pumped into a
``dark state'', a coherent superposition of the two ground-state
sublevels which is decoupled from the optical fields. Typically, a
buffer gas is included to restrict the motion of the EIT atoms and
thus lengthen the atomic interaction time with the laser
fields~\cite{arimondo'96pra,wynands'97,helm'01}; and also to
pressure broaden the electronic excited state so as to allow
co-propagating probe and control fields that interact with most of
the atoms despite Doppler broadening~\cite{javan'02,lee'03}.
As shown in Fig.~\ref{setup.fig}, we employed an external cavity
diode laser (ECDL) tuned to the $\mathrm{D}_1$ line of
${}^{87}$Rb. The total available laser power was about $13$~mW.

In this regime, atomic diffusion is important, yet rather subtle
to model effectively. To date EIT models have treated diffusion
phenomenologically, by assuming a simple homogeneous decay of the
ground state coherence characterized by the timescale $\tau_D$ of
the lowest order diffusion mode across the laser
beam~\cite{happer72,arimondo'96pra,helm'01}. Some recent
experiments have seen indications of the inadequacy of this simple
approach~\cite{zibrov'01ol,zibrovmatsko02pra,novikova05josab}.
Here we report experiments that show that diffusion of atomic
coherence in-and-out of the laser beam (i.e., from the region
illuminated by the optical fields to the surrounding,
unilluminated region, and then back into the laser beam) plays a
crucial role in determining the EIT resonance lineshape and the
lifetime of stored light. As described below, we observed a narrow
central peak in the EIT resonance (much narrower than $1/\tau_D$)
due to the contribution of atoms diffusing in-and-out of the laser
beam. We then eliminated this narrow peak by applying transverse
gradients in the longitudinal magnetic field to decohere atoms
that diffuse well out of the laser beam. Finally, we measured the
influence of atomic coherence diffusion on light pulses that have
been stored in the atomic ensemble using a dynamic EIT technique.

\section{Experimental setup}

We used a polarizing beam splitter (PBS) to create a bypass
reference beam.  We modulated the phase of the main laser field at
$6.835$~GHz (near the ${}^{87}$Rb ground-state hyperfine
splitting) using an electro-optical modulator (EOM)
, so that approximately $2\%$ of the total laser power was
transferred to each first order sideband.
%
All three optical fields then passed through an acousto-optical
modulator (AOM), shifting the frequencies of all fields by
$+80$~MHz. We tuned the laser such that the main carrier frequency
field was resonant with the $5S_{1/2} F=2 \rightarrow 5P_{1/2}
F^\prime=2$ transition of ${}^{87}$Rb; in this case the $+1$
sideband was resonant with the $5S_{1/2} F=1 \rightarrow 5P_{1/2}
F^\prime=2$ transition.
We neglected any influence of the
far-detuned $-1$ sideband.
%

Before entering the Rb vapor cell the laser beam was weakly
focused to a $0.8$~mm diameter spot and circularly polarized using
a quarter-wave plate $(\lambda/4)$. We mounted the cylindrical
glass cell, containing isotopically enriched ${}^{87}$Rb and
either $5$~Torr or $100$~Torr Ne buffer gas
($D=35.7~\mathrm{cm^2/s}$ and $D=1.78~\mathrm{cm^2/s}$
respectively~\cite{vanier74pra,franz76}), inside a three-layer
magnetic shield, which reduced stray magnetic fields to less than
$100~\mu\mathrm{G}$ over the interaction region. A solenoid inside
the magnetic shields provided a weak bias magnetic field $B_0$
($\le 100$~mG). We controlled the temperature of the cell to $\pm
0.2$~K using a blown-air oven. For the EIT lineshape measurements
we kept the temperature low enough ($45^\circ$C for the $5$~Torr
Ne cell, and $55^\circ$C for the $100$~Torr Ne cell) to ensure an
optically thin Rb vapor; for stored light experiments we operated
with the cell at $65^\circ$C to provide significant pulse delay
and storage.

After traversing the cell, the laser beam was combined with the
bypass reference beam on a non-polarizing beam-splitter (NPBS) and
sent to a fast photodetector (FPD). Since the frequency of the
reference field was $80$~MHz lower than the control field, we
detected the $+1$ sideband by measuring the amplitude of the beat
note at $6.915$~GHz using a microwave spectrum analyzer.

\section{Gradient coils}

Pulsed magnetic field gradients are a well established tool in NMR
for imaging (MRI) and for measuring diffusion and coherent flow in
liquids and gases.
\begin{figure}
\includegraphics[width=0.6\columnwidth]{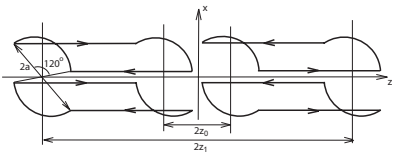}
\caption{ Schematic of the magnetic-field-gradient coils.
Dimensions are: $\mathrm{\textbf{z}}_0= 2.4$~cm,
$\mathrm{\textbf{z}}_1= 16.2$~cm, $\mathrm{\textbf{a}}= 6.4$~cm,
chosen to optimize the linearity of the gradient field $\partial
B_z/\partial x$~\cite{callaghan_book,jin_book}. }
\label{coils.fig}
\end{figure}

In our experiment we used a standard gradient coil design (see
Fig.~\ref{coils.fig})~\cite{callaghan_book,jin_book} to produce a
controllable linear gradient of the longitudinal magnetic field in
the transverse direction: $\partial B_z/\partial x$ up to $40~
\mathrm{mG/(cm\cdot A)}$, with excellent linearity at the center
of the coils.

Note that away from the coil center a transverse magnetic field is
also created such that $\partial B_z/\partial x =
\partial B_x/\partial z$. In traditional high-field NMR
applications, the effect of the transverse magnetic field is
negligible. In our experiment the weak bias magnetic field $B_0$
limits the gradient strength we can use without distorting the EIT
resonance: $\partial B_x/\partial z \cdot L/2 \ll \sqrt{2}B_0$,
where $L$ is the length of the cell.

\section{Steady-state EIT in buffer gas cells}

\begin{figure}
\includegraphics[width=0.6\columnwidth]{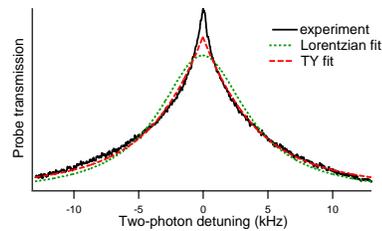}
\caption{Measured intensity of the probe field transmission (EIT)
as a function of two-photon detuning $\delta$. Also shown are the
best Lorentzian fit (\protect{Eq.~(\ref{lorentz})}) and ``TY-fit''
(\protect{Eq.~(\ref{tyfit})}), which assumes spatially
inhomogeneous power broadening of the EIT lineshape as determined
by a Gaussian distribution of the transverse laser intensity
profile. Total laser power $\approx 50~\mu$W. }
\label{lineshape.fig}
\end{figure}

%
Fig.~\ref{lineshape.fig} shows a typical example of the measured
EIT lineshape (i.e. the probe field transmission as a function of
two-photon detuning) for the $5$~Torr Ne cell and no applied
magnetic fields. Note the sharp, narrow peak on-resonance. Also
note that the FWHM is significantly narrower that $1/\pi \tau_D
\approx 26$~kHz~\cite{happer72,arimondo'96pra}, i.e., the
linewidth set naively by single-pass diffusion across the laser
beam.
The shape of this EIT resonance is clearly not the Lorentzian
function predicted by  simple three-level EIT
theory~\cite{lee'03}:
\begin{equation} \label{lorentz}
|\Omega_P^{(out)}|^2 \propto
1-\frac{\gamma_\mathrm{bc}|\Omega_C|^2+
\delta^2\gamma^2}{|\Omega_C|^4+\delta^2\gamma^2}
\end{equation}
where $\Omega_P$ and $\Omega_C$ are the probe and control field
Rabi frequencies, $|\Omega_P^{(out)}|^2$ is proportional to the
probe field transmission intensity, $\delta$ is the two-photon
detuning, and $\gamma$ and $\gamma_{bc}$ are respectively the
excited-state and ground-state coherence relaxation rates (see
Fig.~\ref{setup.fig}a).

Recent extensions of three-level EIT theory have considered the
effect of atomic motion in a laser beam with a Gaussian intensity
profile in the transverse direction.
In the regime of high buffer gas pressure and moderately high
laser intensity,  the motion of alkali atoms across the laser beam
is assumed to be slower than the excited and ground-state
relaxation rates, and the atomic coherence is assumed to adjust
instantaneously to the light intensity at each point of the laser
beam. Thus atoms in the center of the laser beam, where the laser
intensity is maximum, have greater power broadening than atoms in
the wings. For a Gaussian laser intensity profile, the probe field
transmission lineshape is then found to
be~\cite{taichenachev'02iqec}:
\begin{equation} \label{tyfit}
|\Omega_P^{(out)}|^2 \propto
1-\frac{\delta\gamma}{|\Omega_C|^2}\arctan{\frac{|\Omega_C|^2}{\delta\gamma}}
\end{equation}
which we refer to as the ``TY-fit''.
In the regime of low buffer gas pressure and very low light
intensity, there is negligible power broadening and the EIT
lineshape is calculated to be~\cite{pfleghaar93}:
\begin{equation} \label{weisfit}
|\Omega_P^{(out)}|^2 \propto e^{-|\delta\cdot t_{\mathrm{tr}}|}
\end{equation}
where $t_{\mathrm{tr}}=2r/\langle v\rangle$ is the average alkali
atom transit time through the laser beam, and $\langle v\rangle$
is the average thermal radial velocity of the alkali atoms.
Interestingly, the lineshapes described by Eqs.~(\ref{tyfit}) and
(\ref{weisfit}) are very similar for typical conditions for Rb
vapor EIT, despite being obtained in very different limits.
In
Fig.~\ref{lineshape.fig} we plot the best Lorentzian fit
(Eq.~(\ref{lorentz})) and TY-fit (Eq.~(\ref{tyfit})), which is
more relevant to our experimental conditions. The Lorentzian fit
provides a poor approximation to our measurements; the TY-fit is
superior to the Lorentzian fit, but it fails to reproduce the
sharp structure on resonance.

The observed EIT lineshape may be understood if we assume that the
diffusion of Rb atoms in-and-out of the laser beam creates a broad
range of interaction times rather than a single average diffusion
time $\tau_D$. That is, we must account not only for atoms that
diffuse once through the laser beam and then decohere (never
return), but also those that diffuse out of the laser beam and
return, and thus interact with the laser fields multiple times.
For these returning atoms the interaction with the optical fields
resembles a Ramsey separated-oscillatory-field
experiment,
with significant coherent evolution ``in-the-dark'', leading to
much narrower EIT resonances.

\begin{figure}
\includegraphics[width=0.9\columnwidth]{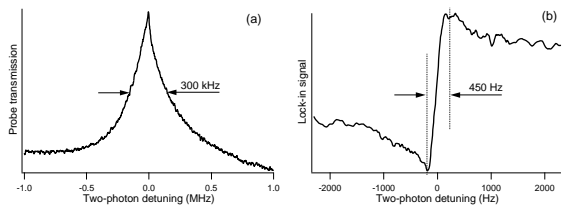}
\caption{EIT resonance observed (a) with spectrum analyzer and (b)
using phase-sensitive detection. Both measurements are made in a
$5$~Torr Ne cell at total laser power of $800~\mu$W.}
\label{lockin_sample.fig}
\end{figure}

To better characterize the sharp central peak of the EIT
resonance, we employed phase-sensitive detection. We used
audio-frequency-modulation of the $6.8$~GHz synthesizer
($f_m\approx 150$~Hz), and detected changes in the laser
transmission using a slow photodetector and lock-in amplifier.
Fig.~\ref{lockin_sample.fig} shows EIT resonance measurements made
both with a spectrum analyzer (covering the entire resonance) and
phase-sensitive detection (restricted to the narrow center of the
resonance). For both the $5$~Torr and $100$~Torr Ne cell, we
measured the spectral width of the narrow central EIT peak to be
$\le 500$~Hz, and to be largely insensitive to laser power, which
is consistent with the central peak reflecting coherent atomic
evolution ``in-the-dark''.
\begin{figure}
\includegraphics[width=0.9\columnwidth]{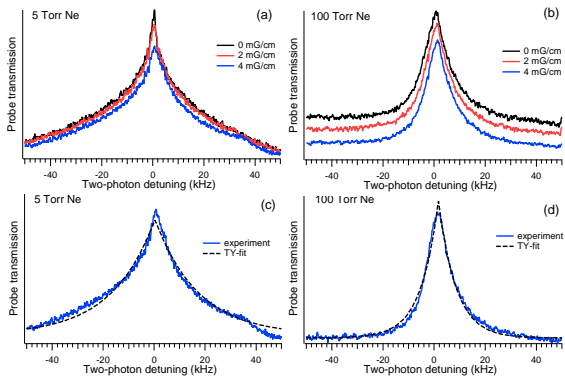}
\caption{Modification of EIT resonance with applied magnetic field
gradients for (a) $5$~Torr Ne cell  and (b) $100$~Torr Ne cell.
Bias magnetic field $B_0\approx 132$~mG, total laser power is
$\approx 100~\mu$W. Comparison of measured EIT resonance for
$4$~mG/cm gradient with TY-fit (\protect{Eq.~(\ref{tyfit})}) for
(c) $5$~Torr Ne cell and (d) $100$~Torr Ne cell.}
\label{grad_eit.fig}
\end{figure}
\begin{figure}
\includegraphics[width=0.6\columnwidth]{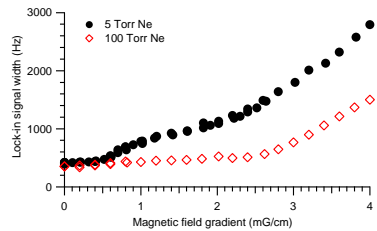}
\caption{Broadening of the EIT central peak with applied magnetic
field gradient measured with phase-sensitive detection, for both
the $5$~Torr Ne and the $100$~Torr Ne buffer gas cells. Bias
magnetic field $B_0\approx 80$~mG, total laser power $\approx
20~\mu$W.} \label{lockin_grad.fig}
\end{figure}
To test the hypothesis that atoms diffusing in and out of the
laser beam contribute significantly to the EIT lineshape
(particularly the sharp central peak), we measured the EIT
resonance in the presence of a linear transverse gradient in the
longitudinal magnetic field $\partial B_z/\partial x$, using the
gradient coils described above. For these measurements we also
applied a uniform longitudinal field ($B_0=80$~mG) which splits
the Rb ground-state Zeeman sublevels, and makes negligible the
effects of the transverse component of the magnetic field created
by the gradient coils (for $\partial B_z/\partial x \le 4$~mG/cm).
We studied EIT formed on the ground-state $m=1$ Zeeman sublevels
(i.e., coherence between the $F=1, m_F=1$ and $F=2, m_F=1$
levels). The associated transition frequency between those levels
has a gyromagnetic ratio $\approx 1.4$~MHz/G.
Therefore, $\partial B_z/\partial x$
induces an inhomogeneous broadening to the EIT coherence; and for
suitable magnitudes of this gradient, the broadening across the
laser beam waist ($\approx 0.8$~mm) can be small compared to the
narrow central peak, while the larger magnetic field deviation
outside the laser beam causes most atoms to decohere if they
diffuse out of and then back in the laser field.

Fig.~\ref{grad_eit.fig} shows measured EIT resonances at several
gradient strengths. For stronger gradients, the central peak
becomes less sharp, and the TY-fit of the EIT lineshape is
improved, which is consistent with diffusion-induced
Ramsey-narrowing being destroyed by gradient-field-induced
decoherence of atoms that diffuse out of the laser beam.

Fig.~\ref{lockin_grad.fig} illustrates the greater dependence on
applied gradient of the spectral width of the central EIT peak for
the $5$~Torr Ne cell, in comparison to the $100$~Torr Ne cell.
These measurements are also consistent with diffusion-induced
Ramsey-narrowing, which would be more thoroughly inhibited by the
applied magnetic field gradient for the more rapid Rb diffusion in
the $5$~Torr Ne cell. Modelling and quantitative analysis of these
diffusion effects will be reported in a future publication.

\section{Stored light in a buffer gas cell}

Under EIT conditions, the strength of the control field determines
the coherent coupling between the probe field and the ground-state
atomic coherence and thus the group velocity of probe pulse
propagation through the medium~\cite{boydPiO2002}. Appropriate
variation of the control field allows storage of the probe pulse
in the form of an atomic ensemble
coherence~\cite{fleischhauer02pra,lukin03rmp}. In realistic atomic
systems, the maximum storage time is determined by the atomic
coherence lifetime, including the effect of coherence diffusion.
%

\begin{figure}
\includegraphics[width=0.9\columnwidth]{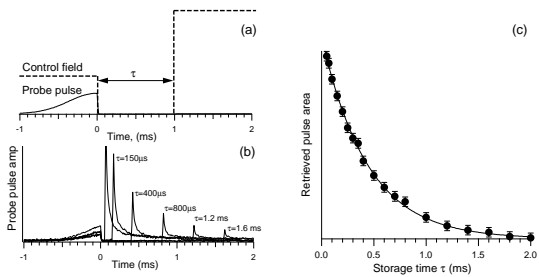}
\caption{(a) Schematic of the timing and amplitude used for the
probe pulse and dynamic control field in measurements of light
storage as a function of storage interval $\tau$. The control
field power is $50~\mu\mathrm{W}$ during the pulse entry stage and
$600~\mu\mathrm{W}$ during the pulse retrieval stage. (b) Examples
of measured probe pulse storage for various storage intervals. (c)
Retrieved pulse area as a function of storage time $\tau$. The
fitted $1/e$ decay time (solid line) is about $500~\mu\mathrm{s}$.
} \label{storedlight.fig}
\end{figure}

To avoid large additional absorption and pulse reshaping, the
bandwidth of the probe pulse should be less than the the sharp
peak observed in the static probe field
transmission~\cite{SPIEproc05}. We employed a Gaussian waveform
for the probe pulse with a full width of $1$~ms, which experienced
a group delay $\approx 450~\mu$s for a control field power of
$50~\mu$W. This delay is not great enough to trap the entire probe
pulse inside the atomic vapor cell ($L = 7$~cm). We typically
stored about half the Gaussian pulse, as shown in
Fig.~\ref{storedlight.fig}(a). Note that we used a larger control
field intensity  at the retrieving stage ($600~\mu$W) to increase
the EIT width of the atomic medium and thereby minimize losses
during release and propagation of the retrieved pulse.

The measured amplitude of the probe pulse for various storage
intervals is shown in Fig.~\ref{storedlight.fig}(b). The $1/e$
decay time of the retrieved pulse area is $\approx 500~\mu$s (see
Fig.~\ref{storedlight.fig}(c)), corresponding to a decoherence
rate $\approx 600$~Hz, in agreement with the width of the sharp
central peak of the EIT resonances at low light intensity, and
much narrower than the rate $1/\pi \tau_D \approx 26$~kHz
associated with diffusion of Rb atoms out of the laser beam.

%
%
\begin{figure}
\includegraphics[width=0.9\columnwidth]{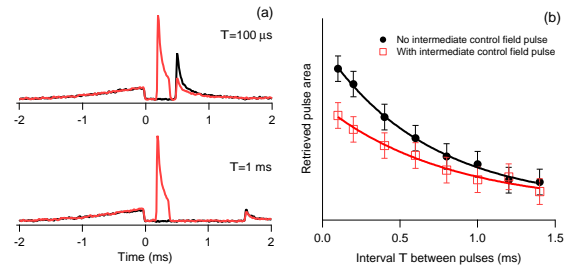}
\caption{(a) Comparison between retrieved light pulses with and
without an intermediate control field pulse for
\textsf{T}$=100~\mu$s and \textsf{T}$=1$~ms.
   Control field power is $50~\mu\mathrm{W}$ at the writing stage
   and $600~\mu\mathrm{W}$ for both retrieving pulses.
    (b) Area of the retrieved probe
    pulses  with (circles) and without (squares) an intermediate
    control field pulse.}
\label{double_readout.fig}
\end{figure}
The large spatial distribution of atomic coherence stored outside
the laser beam  allows multiple retrieved pulses to be observed by
turning the control field on and off several times during the
retrieval process. As a demonstration, we stored a signal pulse in
the atomic ensemble; turned the control field off for $200~\mu$s;
then on for another $200~\mu$s to retrieve a first pulse; next
turned off the control field again for a variable time \textsf{T};
and finally turned the control field on and retrieved a second
pulse. Fig.~\ref{double_readout.fig} shows the results of these
measurements and a comparison to experiments with no intermediate
control field pulse.
For small \textsf{T}, there is a large difference between the
retrieved pulse areas with and without an intermediate control
field application, since the second application of the control
field mostly interacts with the same atoms from which coherence
has already been retrieved by the first pulse. For larger values
of \textsf{T}, however, atomic coherence diffuses back into the
laser beam, and the difference between the pulses retrieved with
and without an intermediate control field pulse decreases.

\section{Conclusions}

We have presented an initial experimental study of the effect of
atomic coherence diffusion in-and-out of the laser beam on EIT and
light storage in a Rb vapor cell with Ne buffer gas. We found that
the EIT resonance lineshape is not well described by models that
account for atomic diffusion with a simple homogeneous decay rate
of the ground state. We also observed a sharp, narrow peak of the
EIT resonance that can be understood qualitatively due to the
coherent atomic evolution ``in-the-dark'' as atoms diffuse
in-and-out of the laser beam. We found that this diffusion-induced
Ramsey-narrowing could be reduced by application of a suitable
magnetic field gradient, $\partial B_z/\partial x$, which
decoheres most atoms that diffuse out of the laser beam. We also
measured light storage times consistent with the inverse of the
linewidth of the narrow central peak of the EIT resonance, and
observed ``replenishment'' of stored light signals due to atomic
coherence diffusion from the region outside of the laser beam.

The authors are grateful to F.\ Can\'{e}, M.\ Klein, C.\
Smallwood, M.\ D.\ Lukin, M.\ D.\ Eisaman, M.\ Bajcsy, L.\
Childress, A.\ Andr\'{e}, M.\ Crescimanno, A.\ Weis, and A.\ S.\
Zibrov for useful discussions, and to Christine Y.-T. Wang for
construction of the gradient coils. This work was supported by
DARPA, ONR and the Smithsonian Institution.

\end{document}